\documentclass[preprint,12pt]{elsarticle}

\usepackage{amsmath,amssymb}
\journal{Journal of Magnetism and Magnetic Materials}

\begin{document}

\begin{frontmatter}

\title{Anomalous Hall Effect in Bismuth}

\author[aff_1]{Bruno Cury Camargo\corref{mycorrespondingauthor)}}
\cortext[mycorrespondingauthor]{Corresponding author. Tel. +48-22-1163334}
\ead{b.c\_camargo@yahoo.com.br}
\author[aff_1]{Piotr Gier{\l}owski}
\author[aff_2]{Andrei Alaferdov}
\author[aff_3]{Iraida N. Demchenko}
\author[aff_1]{Maciej Sawicki}
\author[aff_1]{Katarzyna Gas}
\author[aff_2]{Yakov Kopelevich}

\address[aff_1]{Institute of Physics, Polish Academy of Sciences, Aleja Lotnikow 32/46, PL-02-668 Warsaw, Poland.}
\address[aff_2]{Instituto de Fisica Gleb Wattaghin, R. Sergio Buarque de Holanda 777, 13083-859 Campinas, Brazil.}
\address[aff_3]{University of Warsaw, Department of Chemistry, ul. Pasteura 1, PL-02-093 Warsaw, Poland.}

\begin{abstract}
We report the occurrence of ferromagnetic-like anomalous Hall effect (AHE) below $30$ mT in bismuth crystals.  The signatures of ferromagnetism in transport are not corroborated in magnetization measurements, thus suggesting the induction of non-intrinsic magnetism at surfaces and grain boundaries in bismuth. The suppression of the AHE with the increase of magnetic field and temperature coincides with previous reports of superconductivity in Bi, suggesting an interplay between the two phenomena.
\end{abstract}

\end{frontmatter}
%\linenumbers

\section{Introduction}

Bismuth (Bi) is perhaps the original wonder material, being the catalyzer for the discovery of game-changing phenomena in condensed matter physics – such as the Shubnikov-de Haas, de Haas-van Alphen, Seeback and Nernst effects \cite{Shubnikov1930, deHaas1930, Ettingshausen1886}.  More recently, it has also been shown to host intrinsic superconductivity, to behave as a higher order topological insulator and predicted to be a topological crystalline insulator - a testament to its unusual properties \cite{Prakash2017, Schindler2018, Hsu2019}.

Being a material with low charge carrier concentration ($\approx 10^{17}$ $\text{cm}^{-3}$), low carrier effective masses ($\approx 10^{-3}$ $\text{m}_e$) and a small density of states at Fermi level ($N(0) \approx 0.1 \text{ eV}^{-1}$) \cite{Mikhail1981, Edelman1976, Madelung_EMB, Madelung_BS, Orovets2012}, Bi is susceptible to Fermi surface (FS) reconstructions, which manifest as exotic electronic states.  These can be triggered by defects or external parameters, such as magnetic field and pressure \cite{Kopelevich2006, Koley2017, Armitage2010}. It is known, for example, that Bi undergoes a metal-to-insulator transition when exposed to magnetic fields of the order of few Tesla - with interpretations including the coexistence of electrons and holes within the two-band model \cite{Du2005}, the development of an excitonic gap and the occurrence of magetic-field-induced electron-electron pairing \cite{Kopelevich2006}.

Historically, magnetotransport properties of bismuth have been studied at magnetic fields above $1$ T, partially due to intriguing properties such as linear magnetoresistance, strong spin-orbit coupling and the low magnetic fields necessary to attain the quantum limit \cite{Behnia2007, Kapitza1928}. Here, we focus on the low magnetic field magnetoresistance. We demonstrate the occurrence of hysteresis loops in Hall measurements consistent with the presence of magnetic ordering in Bi which, being one of the most diamagnetic non-superconducting materials known, is not expected to exhibit either ferro- or antiferromagnetism.

\section{Samples and experimental details}

We investigate crystalline Bi samples cleaved from large Bi crystals grown by the Brigdman method, with 5N purity (see the Supplementary information for images, surface and chemical characterization). The specimens were shaped as rectangular bars with approx. dimensions $5 \text{ mm} \times 3 \text{ mm}$ (in-plane) and thickness between $0.5$ and $1$~mm. Surfaces exhibited many grain boundaries and quasi-decoupled crystallites. Experiments comprised of electrical resistivity and magnetization measurements at temperatures $2 \text{ K} \leq T \leq 10 \text{ K}$ and magnetic fields $-30 \text{ mT} \leq \mu_0 H \leq 30 \text{ mT}$. The magnetic fields were generated by superconducting coils, and the chosen field range ensured a fixed remnant field throughout experiments.

Magnetization measurements were carried out in a MPMS XL Superconducting Quantum Interference Device (SQUID) magnetometer equipped with a low field option. A soft quench of the superconducting magnet was executed prior the measurements to assure a below 0.1~mT magnitude of the trapped field in the sample chamber. Long Si strips facilitated the samples support in the magnetometer chamber and we strictly followed the experimental code and data reduction detailed in Ref.~\cite{Sawicki_2011}.

Transport measurements were independently carried in a homemade $\text{He}_4$ cryostat and a Quantum Design 9T PPMS instrument, yielding similar results. Samples were contacted in the standard five-probe geometry and the five-probe Hall geometry using gold wires either directly soldered to the sample surface or glued with non-superconducting silver epoxy. In total, seven samples were measured, labeled Bi1 to Bi7. Devices Bi1 to Bi5 were prepared in the five-probe geometry and underwent simultaneous longitudinal and transverse resistivity measurements both with DC and AC excitation currents. Samples Bi6 and Bi7 were contacted in the five-probe Hall geometry and only had their transverse resistivity characterized. Details about contact configuration are presented in the supplementary information. Experiments in the presence of magnetic fields occurred in zero-field-cooled (ZFC) and field-cooled (FC) regimes. Between measurements, samples were systematically demagnetized at $T = 10$ K, from a field of $1$ T to ensure consistent remnant fields in the superconducting coil. In what follows, we present results obtained for samples Bi1 and Bi6. Additional samples are available in the supplementary material.

\section{Results and discussion}

\begin{figure}[b!]
\begin{center}
\includegraphics[width = 8.0cm]{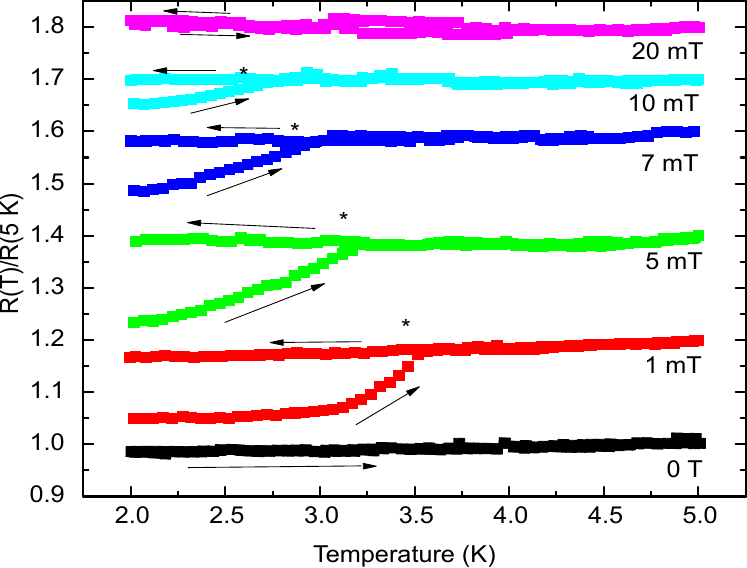}
\caption{Measured $R_{xx}$($T$) for sample Bi1. The curves have been displaced vertically for clarity. The direction of the measurements are indicated by arrows. Measurements at $0$ and $20$ mT presented reversible behavior. Other curves presented an irreversibility between heating and cooling cycles, indicated by stars.}
\label{fig_RvsT}
\end{center}
\end{figure}

The main result of the present report is shown in Fig. \ref{fig_RvsT}. Electrical transport measurements in our devices revealed featureless $R$($T$) curves in the absence of magnetic fields. For small values of $\mu_0H$, however, a sharp (up to $25$\%) resistance reduction was observed below 4~K. The transition was suppressed by magnetic fields, surviving up to $\mu_0 H \approx 20$ mT.% Above such value, the featureless behavior was recovered.

The presence of this transition was strongly dependent on the samples magnetothermal history. It was triggered below certain temperatures $T^*$, associated to an irreversible behavior in the $R$($T$) curves (see Fig. \ref{fig_RvsT}). Resistance drops were observable only in ZFC measurements, whereas FC curves were featureless below $5$~K. % The usance of magnetic fields yielded definite efficaciousness upon the transmogrification of the transition under scrutiny, diminishing its temperature. This is shown in fig. \ref{fig_RvsT}. 
%, being suppressed by magnetic fields

$R$($T$) measurements performed with different temperature sweeping rates and at thermal equilibrium yielded identical results. Similarly, no frequency-dependency was present on AC resistivity up to $1$ kHz. $I$--$V$ characteristics in the ZFC and FC branches of the $R$($T$) curves revealed ohmic behavior, and no relaxation was observed at $T = 2$ K in measurements spanning eight hours. These results weigh against a charge density wave as the source of the hysteresis observed. They do not discard, however, the possibility of a glassy state with typical relaxation times above several hours.

Instead, the presence of irreversibility in $R$($T$) measurements can be attributed to the existence of small, randomly-aligned regions with single magnetic domains within the material.%, as large regions would be more likely to host domain walls and therefore produce loops with larger ZFC resistances than those in the FC regime.
 In the ZFC condition, the pinning of domains at small applied $\mu_0 H$ ensures a sample with weaker macroscopic magnetic response at lower temperatures. In the FC regime, however, the cooling in the presence of $\mu_0 H \neq 0$ causes the same effective field above and below $T^*$, suppressing the transition.%landscape with smaller effective magnetic fields

\begin{figure}
\begin{center}
\includegraphics[width = 8.0cm]{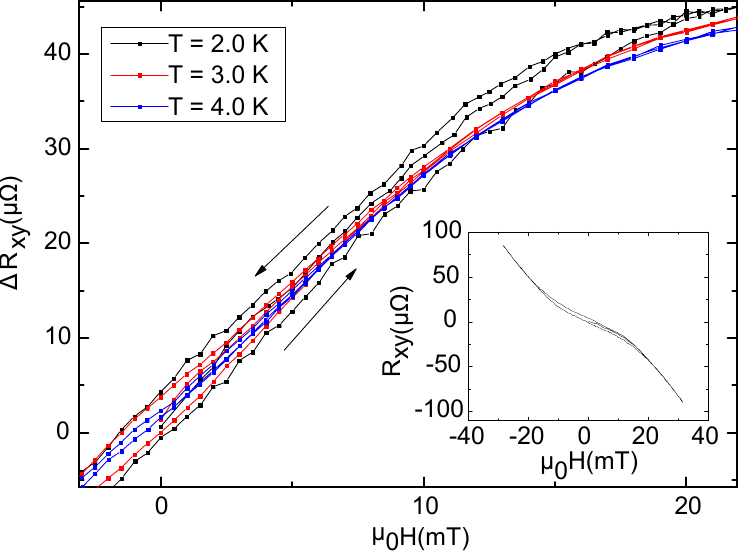}
\caption{Hall component $\text{R}_\text{xy}$ for sample Bi6 at different $T$ after the subtraction of a linear background, $\Delta \text{R}_{xy} \equiv \text{R}_{xy}(H)-\beta H$ with $\beta\approx -43$~$m \Omega$/T. A behavior resembling a ferromagnetic hysteresis developed below $4$ K. The inset shows the raw $R$($H$) curve at $T = 2$ K.}
\label{fig_Hall}
\end{center}
\end{figure}

Such a hypothesis is corroborated by magnetoresistance measurements below $T^*$, which revealed a hysteretic behavior compatible with the presence of pinned magnetic domains. The hysteresis observed in the antisymmetric component of $\text{R}_\text{xy}$, shown in Fig. \ref{fig_Hall} for sample Bi6, resembled those found in the anomalous Hall component of ferromagnets. The temperature dependence of the extracted coercive field was well described by
\begin{equation}
\mu_0 H = 0.96\times\frac{K_1}{M_S}\left[1-\left(\frac{T}{T_0}\right)^{\frac{3}{4}}\right],
\label{eq_kneller}
\end{equation}
with $K_1/M_s \approx 3.6 \times 10^{-3}$ T the ratio between the anisotropy density of the magnetic centers $K_1$, and their magnetization $M_S$. For our samples, $T_0 \approx 4.0$ K, which is close to the critical temperature obtained by mapping $T^*(H)$ %the irreversible behavior of the $R$($T$) curves
 (yielding $T_0 \approx 3.9$ K). A diagram with the results  for sample Bi1 is presented in Fig. \ref{fig_diagram}. A diagram containing results of all measured samples is shown in the supplementary information. Similar behavior has not been observed previously by us in other high magnetoresistance materials (e.g. graphite \cite{Camargo2016}) using the same instrumentation, discarding an instrumental origin for our observations.
%\subsection{}
%\subsubsection{}

\begin{figure}
\begin{center}
\includegraphics[width = 8.5cm]{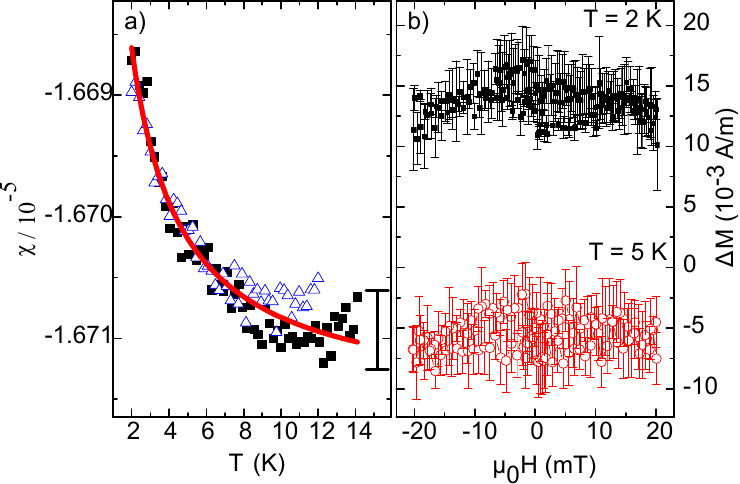}
\caption{a) ZFC (filled squares) and FC (open triangles) $\chi$($T$) curves measured at $\mu_0 H = 10$ mT, exhibiting a weak Curie-Weiss paramagnetic contribution $C/(T+T_0)$, $C \simeq 7.2 \times 10^{-3}$ K and $T_0 \simeq 0.6$~K (red line). The vertical bar corresponds to the experimental uncertainty. b) Magnetization hysteresis loops in Bi measured at $T = 2$ K (closed black symbols) and $T = 5$ K (open red symbols) after the subtraction of a linear diamagnetic background $\Delta M(H) = M(H) - \alpha H$. Curves are shifted vertically for clarity.}
\label{fig_MxT}
\end{center}
\end{figure}

The relation expressed in Eq.~\ref{eq_kneller} corresponds to the coercivity of a non-interacting assembly of single domain magnetic particles \cite{Kechrakos1998} with blocking temperature $T_0$. The exponent $3/4$ corresponds to the case when their easy axis is randomly-oriented with respect to the magnetic field \cite{Battle1993}. Such an observation in Hall measurements suggests that our samples present weak, diluted magnetism. Magnetization measurements, however, yielded no signs of ferromagnetic-like behavior. % within experimental resolution.
 The $M$($T$) dependency was diamagnetic with a weak Curie-Weiss paramagnetic signal at the lowest $T$, Fig.~\ref{fig_MxT}~a). Neither the nonlinear component of the diamagnetic-dominated $M$($H$) data (see the Suppl. Material, Fig. S6), shown in Fig.~\ref{fig_MxT}~b), points towards ferromagnetic-like behavior. Rather, it can be attributed to weak irreproducibilities of the magnet power supply \cite{Gas2019}.
 Combined, these observations further indicate a very small volumetric fraction as the responsible for the phenomenon found in electrical transport measurements.% \ms{something is missing here, I think.}

Indeed, elemental analysis of the samples did not reveal the presence of magnetic impurities (see the Suppl. Material). Combined to magnetization measurements, these results attest against contaminants as the source of the AHE. For example, taking the saturation magnetization of iron \cite{Crangle1971} at $M_{sat} \approx 2.73 \times 10^{-4} \text{ Am}^2/\text{kg}$, the absence of signatures of magnetism down to $10^{-10}$ $\text{Am}^2$ ($9 \times 10^{-7} \text{Am}^2/\text{kg}$) puts an upper limit for Fe impurities at $5$ $\mu$g in our $0.111$ g sample, or $45$ parts per million in mass. For magnetite ($\text{Fe}_2\text{O}_3$, $M_{sat} \approx 9.42 \times 10^{-5} \text{ Am}^2/\text{kg}$), a similar estimation yields $120$ ppm in mass\cite{Hu2017}.

\begin{figure}
\begin{center}
\includegraphics[width = 8.0cm]{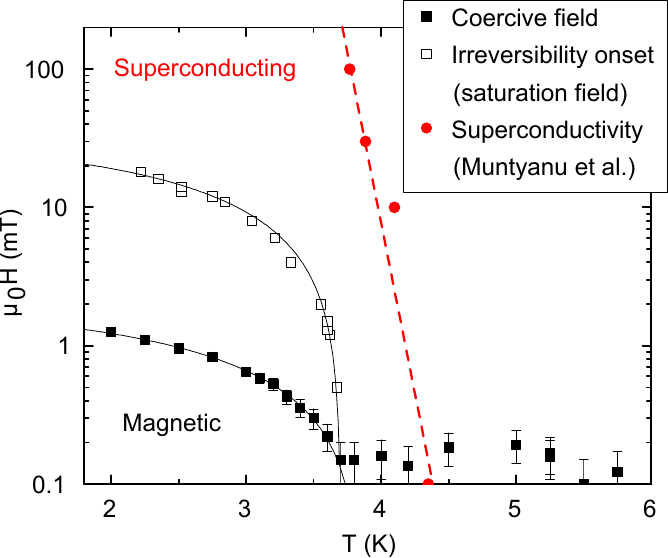}
\caption{Magnetic field vs. temperature diagrams for the coercive field (black closed squares) and irreversibility temperature $T^*$ (open squares) of the magnetic phase of Bi. Results correspond to sample Bi1.  Red circles represent the superconducting phase diagram of bismuth, extracted from Ref. \cite{Muntyanu2004}. The dashed line is a guide to the eye. The full lines represent Eq. \ref{eq_kneller} with $T_0 = 4.0$ K and $K_1/M_s = 3.6$ mT for the closed squares and $T_0 = 3.9$ K and $K_1/M_s = 50$ mT for the open squares.}
\label{fig_diagram}
\end{center}
\end{figure}

Considering the larger value estimated above, the observed magnetic signal in our devices would originate from, at most, $200$ ppm of the total sample volume (ca. $2.3 \times 10^{-12} \text{ m}^3$, whereas the entire sample has approx. $10^{-8} \text{ m}^3$). Taking the noise level of $M$($H$) measurements ($\approx 10^{-10} \text{ Am}^2$) as the upper limit for the saturation magnetic moment of magnetite nano-precipitates yields an estimation for the magnetic centers magnetization $M_S\approx 44$ A/m. From the experimental ratio $K_1/M_S \approx 3.6 \text{ mT}$ (see Eq. \ref{eq_kneller}), we obtain $K_1 \approx 0.154 \text{ J/m}^3$. This value is unreasonably small (typical values for $K_1$ range between $10^4$ and $10^5 \text{ J/m}^3$, and can average only to no less then $10^2 \text{ J/m}^3$ in soft nanocrystalline magnets \cite{Kechrakos1998, Kai1993}), allowing for a re-estimation of the magnetic volumetric fraction of our samples at least three orders of magnitude below $200$ ppm. % (which would increase $M_S$, and therefore, $K_1$ accordingly).
 This new number is consistent with the weak paramagnetic (PM) background in $M$($T$) measurements, which yields an estimated concentration of PM $ = 1.4 \times 10^{15}$ $S = 2$ spins per gram in the sample. Assuming one spin per foreign atom and taking into account the molar mass of Bi at  83~g/mol, this results in a total atomic impurity amount of the order $0.2$~ppm.

Such a small fraction cannot be held accountable for the $\approx 20$\% variation observed in $R$($T$) and the hysteresis loops found in Hall measurements. The transition temperatures found in the experiments % to the best of our knowledge,
 also do not coincide with the typical Curie temperatures of common magnetic contaminants. One way to reconciliate our findings in $M$($T$) and $R$($T$), is to consider the AHE originating at the surface or grain boundaries of Bi, rather than caused by diluted contaminants.

Surely, it has been demonstrated that surface conductivity in bismuth plays a large role in the macroscopic sample resistivity \cite{Hirahara2007}. In particular, it was shown that ad-atoms, dislocations and rotations between crystalline regions can cause the reconstruction of the material’s FS, inducing unusual metallicity and exotic electronic states such as superconductivity \cite{Weitzel1991, Gitsu1992, Muntyanu2004}. Additionally, recent reports revealed that hinge states in Bi along the binary direction, as well as selected crystal surfaces house topological states \cite{Schindler2018, Hsu2019}. The manifestation of the spin-Hall effect in the absence of detectable signs of magnetism, could be held accountable for triggering the AHE observed here.

However, measurements performed before and after in-situ heat treatments in vacuum ($10^{-3}$ Torr at $T = 370$~K for $4$h) yielded no observable changes. This points against a purely adsorption-related phenomenon, as doping caused by adsorbed gases would require a variation of the magnetic properties with the coating, as seen - for example - in graphite \cite{Boukhvalov2011}. This result indicates grain boundaries as the most likely candidate to exhibit magnetism in the system. These occur in all samples, most evidently in cleaved surfaces (see the suppl. information).

These same regions have been linked to superconductivity \cite{Gitsu1992, Muntyanu2006} in Bi with $T_c\approx 4.1$~K. The phenomenon, however, is not intrinsic, but rather a consequence of the reconstructed FS along grain boundaries. It is expected to happen either on amorphous Bi or on interface regions within the sample \cite{Gitsu1992, Muntyanu2006, Hamada1981}. In particular, angle-resolved measurements of quantum oscillations in selected Bi samples have demonstrated \cite{Muntyanu2004} that such regions present a three orders of magnitude larger density of states in comparison with crystalline bulk Bi. This suggests a FS reconstruction with wide band dispersion along dislocations. Similar conclusions have been independently reached much earlier through Hall effect measurements in Bi films (see, e.g. Ref. \cite{Buxo1980} and references therein), with grain boundaries being proposed to host charge carrier concentrations as high as $10^{12}$ $\text{cm}^{-2}$ . This value is four orders of magnitude above the normalized $2D$ charge carrier concentration of bulk Bi, $n_{2D} \equiv n_{3D}\times c_0\approx 10^8$ $\text{cm}^{-2}$, $c_0$ the c-axis lattice parameter ($\approx 0.225$ nm) and $n_{3D}$ the material's bulk charge carrier concentration ($\approx 10^{18}$~$\text{cm}^{-3}$) \cite{Lu1996}. 

While the absence of the Meissner effect and the lack of transitions in FC $R$($T$) measurements attest against the occurrence of superconductivity in our devices, the temperature $T^* = 3.9$ K is indeed close to previously reported $T_c$ - see Fig. \ref{fig_diagram}. We note, however, that superconducting Bi usually presents multiple transitions, around $4.1$ K and $8.4$ K.  Meanwhile, in our samples, no anomaly was observed above $4.1$ K. In addition, the magnetic fields necessary to completely saturate the hysteresis loops in our devices were one order of magnitude below the critical fields reported for superconductivity in bismuth \cite{Muntyanu2006} (see Fig. \ref{fig_diagram}). These observations, combined, weight against superconductivity as the source of magnetism (i.e., the AHE) in our samples. It is possible, however, that both phenomena might be related, with magnetism playing as a precursor of superconductivity, which can be tuned by disorder.

Assuming our sample's magnetic transition as having an intinerant origin with a Curie temperature $T_C \approx 4$K (where the coercive field vanishes),  we estimate the Stoner exchange parameter $I$ that would give rise to magnetism \cite{Kim_book}
\begin{equation}
I\approx \frac{1}{N(0)\left(1-\alpha \left(\frac{k_B T_C}{E_F}\right)^2\right)}\approx 10\text{ eV},
\label{eq_stoner}
\end{equation}
with $E_F \approx 27$ meV the Fermi energy \cite{Mikhail1981}, $N(0) \approx 0.1 \text{ eV}^{-1}$ the density of states of Bi at the Fermi level \cite{Orovets2012} and $\alpha$ a constant below one \cite{Kim_book, Orovets2012}. This value is at least one order  of magnitude above those typical for known ferromagnetic materials (e.g. $I_\text{Fe} \approx 0.6$ eV) \cite{Papaconstantopoulos_book}. This overestimation can be attributed to an inappropriate value of $N(0)$, which is unrealistic in the regions likely responsible for magnetism in our samples – as discussed above. Instead, taking a three orders of magnitude higher density of states at the Fermi level along grain boundaries into account \cite{Muntyanu2004}, we obtain $I \approx 10$ meV.

Different mechanisms could be held accountable for a wide band dispersion resulting in an inflated density of states in Bi. For example, these could be understood in view of natural strains and stresses present in the system. Such features are expected to create an effective vector potential in the material, which under certain conditions can be held accountable for flat or heavy bands \cite{Kauppila2016}. Another possibility arises from the presence of Dirac charge carriers in the bulk and surface of Bi \cite{Schindler2018, Hsu2019}. This type of carriers have been predicted to exhibit universal edge states at crystals boundaries, which can be dispersionless depending on local potentials across the sample surfaces \cite{Shtanko2018}. In our samples, in which crystal grains are grown at random, the presence of strains and/or doped regions satisfying the conditions necessary for the induction of the flat-band states would be satisfied only in a small fraction of the sample. Suitably, the magnetic volumetric fraction previously estimated based on $M$($T,H$) measurements revolves around few ppm. Additionally, different samples grown with the same method show a similar, but non-identical behavior (see sec. D of the SI).

%evices cases can then be associated with the nature of the band dispersion in the system, rather than be caused by scattering. In a spin-Hall system (such as Bi), vector potentials leading to a flattened band could be interpreted as a pseudomagnetic field, breaking the degeneracies responsible for the spin hall effect (SHE) and causing a net magnetization leading to the AHE. A similar scenario would be expected in the case of surface states generated by broken lattice symmetries. In bismuth, these are spin polarized \cite{Schindler2018, Hsu2019}, and 

Flat bands have recently been the focus of intense experimental work in twisted multilayer graphene, in which instabilities towards ferromagnetic and superconducting order are triggered depending on sample doping (see e.g. \cite{Liu2019} and references therein). Similarly, we argue that a  reconstruction of the FS of Bi towards a flat dispersion at crystalline edges can be held accountable for the exotic phenomena observed, with the presence of larger density of states and superconductivity along such regions having already been previously reported \cite{Muntyanu2006}. The triggering of ferromagnetism, however, has not yet been observed.

The Stoner exchange parameter estimated from magnetic measurements is within one order of magnitude of the electron-electron pairing coupling constant $V \approx 3k_B T_c \approx 1.0$ meV calculated in ref. \cite{Kopnin2011} for flat band superconductivity in graphite with $T_c = 4$ K. Despite using an expression obtained for multilayer graphene, we expect its order of magnitude to be appropriate for bismuth as well, as the model employed relies on a layered system with in-plane binding much stronger than interlayer coupling \cite{Kopnin2011}. Although less extreme than in graphite, such a description is suited to bismuth \cite{Ettingshausen1886}. The estimation of superconducting and ferromagnetic exchange interactions around the same order of magnitude suggests a competition between both mechanisms, which can account both for conflicting reports on the properties of bismuth, as well as for the absence of superconductivity in our device.

Unfortunately, we are unable to tune the properties of our devices at will, as done in multilayer graphene \cite{Kauppila2016}. This happens because charge screening prevents the chemical potential in bulk Bi from being varied by gate voltages. Instead, this modulation can be achieved through self-doping imposed by disorder or adsorption of other elements at the sample surface/boundaries. %Hence, different samples are expected to show different properties, explaining potentially diverging reports in the literature about Bi. % In our samples, for example, trace amounts of B (within experimental error) are observed along grain boundaries and the sample surface (see the suppl. Material), which could in principle be held accountable for promoting the necessary doping to trigger magnetism.
Therefore, we propose the interaction between a disordered surface (with grain boundaries and steps) and possibly adsorbed elements to be responsible for the magnetism observed.

Given the purity of our samples, it is unlikely that scattering by magnetic or heavy impurities could be held accountable for the phenomenon reported here. In addition, comparison between magnetization and transport measurements suggests a small fraction of the sample exhibiting magnetism, thus reinforcing the hypothesis that the AHE in our samples do not arise from bulk Bi. At the same time, modeling and experimental evidence suggests that Bi is a non-trivial material whose surfaces exhibit equivalents of the spin-Hall effect (SHE) \cite{Ohtsubo2013, Schindler2018, Hsu2019}, which shares some similarities with the AHE. Considering magnetism having origin along grain boundaries and surfaces, which exhibit larger density of states compared with bulk Bi and are also known to host spin-Hall states \cite{Schindler2018, Hsu2019}, an intrinsic contribution to the AHE seems more likely. In particular, vector potentials triggering a flattened band dispersion in the surface of the material can act as a pseudomagnetic fields \cite{Kauppila2016},  breaking degeneracies responsible for the SHE in the surface of bismuth and causing a net magnetization leading to the AHE.

We close our discussion by mentioning that, while our report might seem surprising at first (experiments are straightforward and magnetic fields are low), results hinting at the AHE in Bi have been previously reported by Conn and Donovan as early as in $1948$ \cite{Donovan1950, Conn1948}. In their results, an anomalous negative MR below $40$ mT was identified in Bi and attributed as due to an anomalous Hall contribution. There, however, no hysteresis measurements were attempted – thus not allowing the attribution of the anomalous MR to magnetism.

\section{Conclusion}

In conclusion, in this work, we demonstrated the occurrence of the AHE in bismuth crystals. The phenomenon appears to be confined to the sample surface or grain boundaries, possibly being triggered by disorder at the interfaces. Circumstantial evidence – such as similar transition temperatures, similar exchange energy parameters and behavior in the presence of magnetic fields - points towards a competition between magnetism and superconductivity, which can be understood in the context of the reconstruction of the FS of Bi along grain boundaries in a flat dispersion. The similarities between results reported here for Bi and those of multilayer graphite are complementary to previous literature and suggest an universal origin for superconductivity in semimetals with low density of states near the Fermi level. We expect additional experiments in different systems to corroborate our hypothesis.

\section*{Acknowledgments}
We would like to thank Marta Cieplak for fruitful discussions. This work was supported by the National Science Center, Poland, research project no. 2016/23/P/ST3/03514 within the POLONEZ programme. The POLONEZ programme has received funding from the European Union's Horizon 2020 research and innovation programme under the Marie Sklodowska-Curie grant agreement No. 665778.  P.G. acknowledges the support of the National Science Center, Poland, research project no. 2014/15/B/ST3/03889. Y. K. and A.A. were supported by CNPq and FAPESP and AFSOR grant  FA9550-17-1-0132

%\bibliography{Database}

\end{document}